\documentclass[10pt, prb, aps, twocolumn, showpacs, citeautoscript, floatfix, reprint, amsmath, amssymb, notitlepage, superscriptaddress]{revtex4-1}

\usepackage{graphicx}
\usepackage{stmaryrd}
\usepackage{rotating}
\usepackage{amsmath}
\usepackage{amsfonts}
\usepackage{amssymb}
\usepackage[countmax]{subfloat}
\usepackage{dcolumn} 
\usepackage{bm} 
\usepackage{color}

\usepackage{float}
\usepackage{latexsym}

\begin{document}

\title{Spin torque and persistent currents caused by percolation of topological surface states}

\author{Wei Chen}

\affiliation{Department of Physics, PUC-Rio, 22451-900 Rio de Janeiro, Brazil}

\date{\today}

\begin{abstract}

The topological insulator/ferromagnetic metal (TI/FMM) bilayer thin films emerged as promising topological surface state-based spintronic devices, most notably in their efficiency of current-induced spin torque. Using a cubic lattice model, we reveal that the surface state Dirac cone of the TI can gradually merge into or be highly intertwined with the FMM bulk bands, and the surface states percolate into the FMM and eventually hybridize with the quantum well states therein. The magnetization can distort the spin-momentum locking of the surface states and yield an asymmetric band structure, which causes a laminar flow of room temperature persistent charge current. Moreover, the proximity to the FMM also promotes a persistent laminar spin current. Through a linear response theory, we elaborate that both the surface state and the FMM bulk bands contribute to the current-induced spin torque, and their real wave functions render the spin torque predominantly field-like, with a magnitude highly influenced by the degree of the percolation of the surface states. On the other hand, impurities can change the spin polarization expected from the Edelstein effect and generate a damping-like torque, and produce a torque even when the magnetization points in-plane and orthogonal to the current direction.

\end{abstract}

\maketitle

\section{Introduction}

A unique feature of three-dimensional (3D) topological insulators (TIs), namely the existence of spin-polarized surface states at low energy, has motivated the search for their applications in spintronic devices. The dispersion of these surface states takes the form of a Dirac cone, with the spin polarization roughly circulating the cone, and the direction of circulation is opposite at energy above and below the Dirac point\cite{Zhang09,Liu10,Yazyev10,Shan10,Zhang12}. Such a spectacular spin-momentum locking profile indicates the possibility of electrically controllable spintronic effects, which can be a great advantage for practical applications. Various recent experiments indeed confirm this type of effects, such as the current-induced spin polarization at the surface of the TI\cite{Tian15,Kondou16,Liu18,Dankert18}. Moreover, the experimentally observed current-induced spin polarization remains roughly constant over a wide range of temperature and chemical potential, which has been attributed to the impurity scattering\cite{Chen20_TI_Edelstein}, signifying the importance of disorder in these surface state-based spintronic effects.



Among the devices that exploit the spintronic effects of the surface states, a particularly promising design that have delivered remarkable performance are the TI/ferromagnetic metal (TI/FMM) bilayers. In particular, the spin pumping experiment in these systems demonstrates their ability to convert the spin current induced by the magnetization dynamics into a charge current\cite{Shiomi14,Jamali15,Wang16_2,RojasSanchez16,Mendes17}. In retrospect, a charge current driven through these systems induces a magnetization dynamics, which may outperform the same phenomenon in the usual heavy metal/FMM heterostructures\cite{Mellnik14,Wang15_spin_torque_TI,Mahendra18}, and has stimulated a great deal of theoretical effort to understand the underlying microscopic mechanisms\cite{Garate10,Yokoyama10,Mahfouzi12,Lu13,Sakai14,Fischer16,Mahfouzi16,Ho17,Ndiaye17,Okuma17_2,Ghosh18,Laref20}.

On the other hand, there are obvious peculiarities regarding the role of the surface states in these spintronic effects in the TI/FMM bilayers. Firstly, the metallic nature of the FMM in TI/FMM bilayers seems to imply that the surface states may no longer be entirely confined in the TI, but extending into the FMM. Secondly, similar to that occurs in two-dimensional metallic materials with Rashba spin-orbit coupling and magnetization\cite{Manchon08,Manchon09,Gambardella11}, the spin-momentum locking profile of the surface states may be altered by the magnetization, which may also modify the spintronic effects of the TI/FMM bilayer. Finally, since the FMM itself certainly contains more conducting channels than the TI, how the bulk bands of the FMM participate in the current-induced spin torque remains to be understood.

In this article, we aim to clarify these issues by means of a lattice model approach. We adopt the philosophy developed recently in a similar system of lower dimension, namely a two-dimensional (2D) square lattice model of TI/FMM side junction\cite{Zegarra20}. The square lattice model delineates the percolation of the edge state of the 2D TI into the FMM, which highly depends on the direction of the magnetization, as well as how the Dirac cone and the FMM bulk bands intertwine. Both the edge states and the bulk bands of the 2D FMM contribute to the current-induced spin torque, and impurities are found to have profound influence on the magnitude of the current-induced spin polarization. In the present work, we advance such a lattice model approach to the 3D TI/FMM bilayers in question. Using a tight-binding model regularized from the low energy sector of the TI/FMM\cite{Ghosh18}, we detail how the magnetization and band structure affect the spin-momentum locking and percolation of the surface states, and unveil a laminar flow of equilibrium persistent charge current controllable by the magnetization. In addition, the bilayer also supports a persistent laminar spin current at equilibrium flowing in both the TI and the FMM. Through a linear response theory that simultaneously takes into account both the surface state Dirac cone and the FMM bulk bands, and without explicitly invoking interface Rashba spin-orbit coupling, we show that the real wave functions of the percolated surface states result in a current-induced spin torque that is predominantly field-like, with a magnitude highly influenced by the percolation. However, the presence of impurities greatly modifies the spin accumulation of the Edelstein effect, and subsequently generates a damping-like spin torque in the FMM. 




\begin{figure}[ht]
\begin{center}
\includegraphics[clip=true,width=0.9\columnwidth]{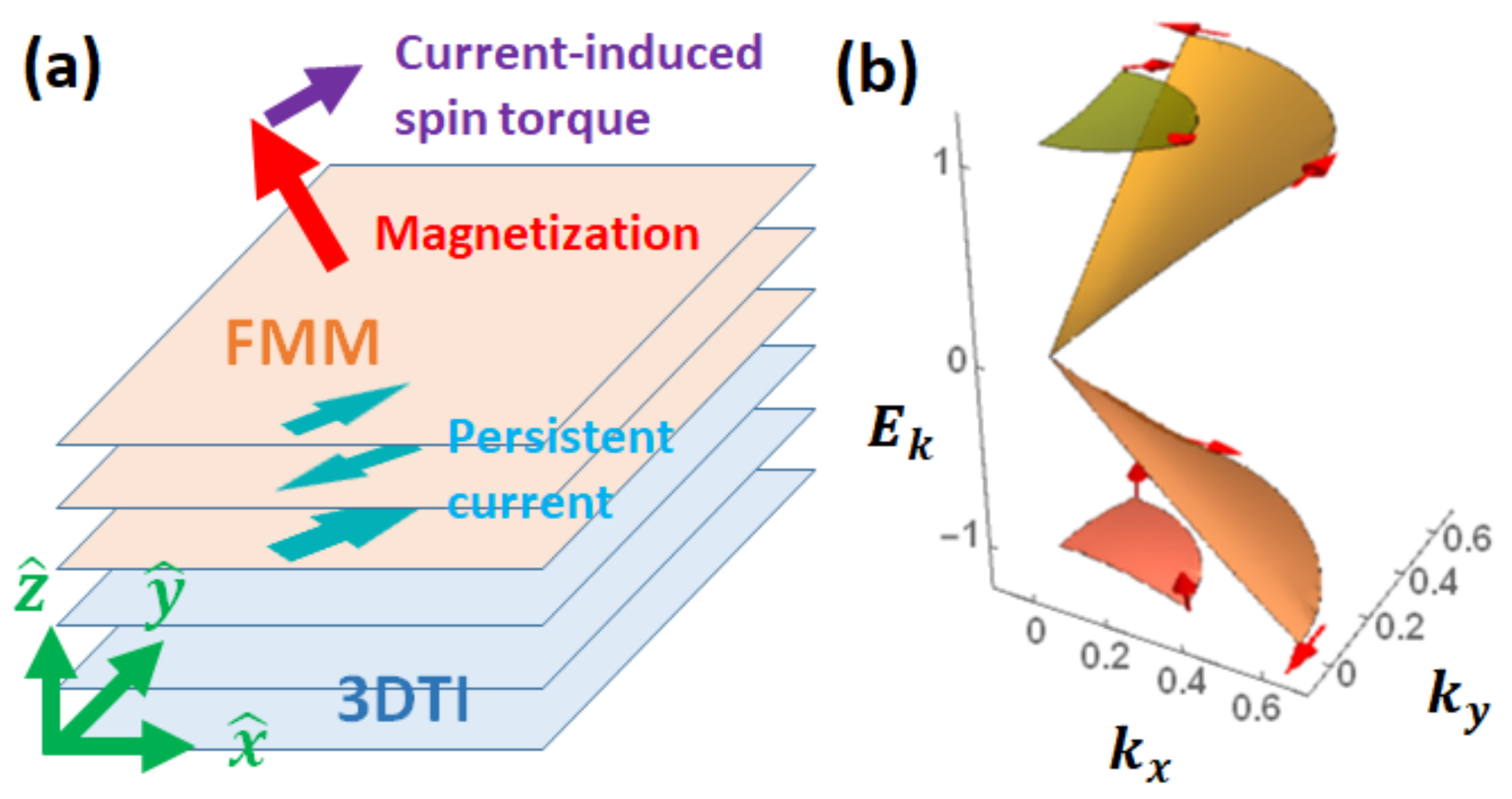}
\caption{(a) Schematics of the TI/FMM slab. (b) The low energy band structure of an isolated TI slab in the first quartet of the BZ. Orange sheets are the surface state Dirac cone, and the red and green sheets are low energy bulk bands. The spin polarization of the eigenstate at few selected points are indicated by red arrows. } 
\label{fig:3DTIFMMslab_schematics}
\end{center}
\end{figure}

\section{TI/FMM bilayers}

\subsection{Constructing the lattice model}

We first discuss the construction of a cubic lattice model a 3D TI thin film, such as Bi$_{2}$Se$_{3}$, deposited on a stack of FMM layers, assuming the stacking direction is along the crystalline ${\hat{\bf z}}$ direction. The low energy sector of the TI is formed by the basis $|P1_{-}^{+},\uparrow\rangle$, $|P2_{+}^{-},\uparrow\rangle$, $|P1_{-}^{+},\downarrow\rangle$, $|P2_{+}^{-},\downarrow\rangle$, where the quantum numbers represent the hybridized Bi and Se orbitals, and the $\left\{\uparrow,\downarrow\right\}$ represents the spin index\cite{Zhang09,Liu10}. We adopt the representation for the $\Gamma$-matrices to construct the Dirac Hamiltonian
\begin{eqnarray}
\Gamma_{i}=\left\{\sigma^{1}\otimes\tau^{1},\sigma^{2}\otimes\tau^{1},\sigma^{3}\otimes\tau^{1},
I_{\sigma}\otimes\tau^{2},I_{\sigma}\otimes\tau^{3}\right\},\;\;\;
\end{eqnarray} 
with the spinor
\begin{eqnarray}
\psi_{\bf k}=\left(
\begin{array}{c}
c_{{\bf k}P1_{-}^{+}\uparrow} \\
c_{{\bf k}P2_{+}^{-}\uparrow} \\
c_{{\bf k}P1_{-}^{+}\downarrow} \\
c_{{\bf k}P2_{+}^{-}\downarrow} 
\end{array}
\right)\equiv
\left(
\begin{array}{c}
c_{{\bf k}s\uparrow} \\
c_{{\bf k}p\uparrow} \\
c_{{\bf k}s\downarrow} \\
c_{{\bf k}p\downarrow} 
\end{array}\right),
\end{eqnarray}
where $s$ and $p$ abbreviate the $P1_{-}^{+}$ and $P2_{+}^{-}$ orbitals, respectively, which are not to be confused with the usual notation of atomic orbitals. The low energy Hamiltonian obtained from ${\bf k\cdot p}$ theory is\cite{Liu10}
\begin{eqnarray}
\hat{H}&=&\left(M+M_{1}k_{z}^{2}+M_{2}k_{x}^{2}+M_{2}k_{y}^{2}\right)\Gamma_{5}
+B_{0}\Gamma_{4}k_{z}
\nonumber \\
&+&A_{0}\left(\Gamma_{1}k_{y}-\Gamma_{2}k_{x}\right)={\bf d}\cdot{\boldsymbol\Gamma}\;,\;\;\;
\label{3D_TI_H0_H1}
\end{eqnarray}
where only lowest order terms essential for the surface states are retained. We construct the lattice model by extending the momentum dependence to the entire Brillouin zone (BZ)
\begin{eqnarray}
k_{\delta}\rightarrow\sin k_{\delta}\delta,\;\;\;k_{\delta}^{2}\rightarrow 2\left(1-\cos k_{\delta}\delta\right),
\end{eqnarray}
where $\delta=\left\{a,b,c\right\}$ are the lattice constants, and then Fourier transform to real space according to
\begin{eqnarray}
&&\sum_{\bf k}\cos{\bf k}\cdot{\boldsymbol\delta}\,c_{{\bf k}A}^{\dag}c_{{\bf k}B}
=\frac{1}{2}\sum_{i}\left\{c_{iA}^{\dag}c_{i+\delta B}+c_{i+\delta A}^{\dag}c_{iB}\right\},
\nonumber \\
&&\sum_{\bf k}i\sin{\bf k}\cdot{\boldsymbol\delta}\,c_{{\bf k}A}^{\dag}c_{{\bf k}B}
=\frac{1}{2}\sum_{i}\left\{c_{iA}^{\dag}c_{i+\delta B}-c_{i+\delta A}^{\dag}c_{iB}\right\},\;\;\;\;
\end{eqnarray}
here $\left\{A,B\right\}$ are combined orbital and spin indices. The FMM is described by the usual quadratic hopping and exchange coupling. Assuming the TI stack has $N_{z,TI}$ layers and the FMM stack has $N_{z,FM}$ layers, we denote
\begin{eqnarray}
&&i\in TI\;\Rightarrow\;z=1,2...N_{z,TI},
\nonumber \\
&&i\in FM\;\Rightarrow\;z=N_{z,TI}+1,N_{z,TI}+2...N_{z,TI}+N_{z,FM}.
\nonumber \\
&&i\in BD\;\Rightarrow\;z=N_{z,TI},
\end{eqnarray}
This leads to our 3DTI/FMM stack cubic lattice model
\begin{eqnarray}
H&=&\sum_{i\in TI,\sigma}\tilde{M}\left\{c_{is\sigma}^{\dag}c_{is\sigma}-c_{ip\sigma}^{\dag}c_{ip\sigma}\right\}
\nonumber \\
&+&\sum_{i\in TI,I}t_{\parallel}\left\{c_{iI\uparrow}^{\dag}c_{i+a\overline{I}\downarrow}
-c_{i+aI\uparrow}^{\dag}c_{i\overline{I}\downarrow}+h.c.\right\}
\nonumber \\
&+&\sum_{i\in TI,I}t_{\parallel}\left\{-ic_{iI\uparrow}^{\dag}c_{i+b\overline{I}\downarrow}
+ic_{i+bI\uparrow}^{\dag}c_{i\overline{I}\downarrow}+h.c.\right\}
\nonumber \\
&+&\sum_{i\in TI,\sigma}t_{\perp}\left\{-c_{is\sigma}^{\dag}c_{i+cp\sigma}+c_{i+cs\sigma}^{\dag}c_{ip\sigma}+h.c.\right\}
\nonumber \\
&-&\sum_{i\in TI,\sigma}M_{1}\left\{c_{is\sigma}^{\dag}c_{i+cs\sigma}-c_{ip\sigma}^{\dag}c_{i+cp\sigma}+h.c.\right\}
\nonumber \\
&-&\sum_{i\in TI,\delta,\sigma}M_{2}\left\{c_{is\sigma}^{\dag}c_{i+\delta s\sigma}-c_{ip\sigma}^{\dag}c_{i+\delta p\sigma}+h.c.\right\}
\nonumber \\
&-&\sum_{i\in FM,\delta I\sigma}t_{F}\left\{c_{iI\sigma}^{\dag}c_{i+\delta I\sigma}+h.c.\right\}
\nonumber \\
&+&\sum_{i\in FM,I\sigma}J_{ex}\,{\bf S}\cdot c_{iI\alpha}^{\dag}{\boldsymbol\sigma}_{\alpha\beta}c_{iI\beta}-\sum_{i\in FM,I\sigma}\mu_{F}c_{iI\sigma}^{\dag}c_{iI\sigma}
\nonumber \\
&-&\sum_{i\in BD,I\sigma}t_{B}\left\{c_{iI\sigma}^{\dag}c_{i+c I\sigma}+h.c\right\},
\label{3DTIFMM_Hamiltonian}
\end{eqnarray}
where $\tilde{M}=M+2M_{1}+4M_{2}$, $I=\left\{s,p\right\}$ and $\overline{I}=\left\{p,s\right\}$ are the orbital indices, $\delta=\left\{a,b,c\right\}$ denotes the lattice constants, $\sigma=\left\{\uparrow,\downarrow\right\}$ is the spin index, and $t_{B}$ is the hopping that controls the interface coupling between the TI and the FMM. The model is schematically shown in Fig.~\ref{fig:3DTIFMMslab_schematics} (a). We will consider the situation that the periodic boundary condition (PBC) is imposed in the planar directions ${\hat{\bf x}}$ and ${\hat{\bf y}}$, and the open boundary condition (OBC) is imposed in the out-of-plane direction ${\hat{\bf z}}$.

The numerical simulation done on a single cluster is constrained by the achievable lattice size of the order of $\sim 10\times 10\times 10$. Thus we choose the following parameters 
\begin{eqnarray}
&&t_{\parallel}=-M=M_{1}=M_{2}=1,\;\;\;t_{\perp}=0.8,
\nonumber \\
&&t_{F}=t_{B}=0.6,\;\;\;J_{ex}=0.1,
\end{eqnarray}
that are order of magnitude similar to that in realistic TIs\cite{Zhang09,Liu10,Ghosh18} and are suitable to draw conclusions from this lattice size. Nevertheless, we emphasize that the statements we obtain is fairly robust against changing of parameters. The FMM chemical potential $\mu_{F}$ controls the two generic types of band structures, as will be discussed in Sec.~\ref{sec:percolation_surface_state}.

Before addressing the TI/FMM bilayers, we first remark on the spintronic properties of the TI alone. Figure \ref{fig:3DTIFMMslab_schematics} (b) shows the low energy band structure of a TI slab of $N_{z,TI}$ layers, equivalent to turning off all the $i\in FM$ and $i\in BD$ terms in Eq.~(\ref{3DTIFMM_Hamiltonian}). The band structure solved by applying a Fourier transform in the planar directions 
\begin{eqnarray}
c_{iI\sigma}=c_{(x,y,z)I\sigma}=\sum_{k_{x},k_{y}}e^{ik_{x}x+ik_{y}y}c_{(k_{x},k_{y},z)I\sigma},
\label{Fourier_transform_kxky}
\end{eqnarray}
clearly captures the Dirac cones of the surface states localized at the two surfaces $z=1$ and $z=N_{z,TI}$, with the Dirac point located at zero energy. Focusing on the Dirac cone of the surface state at the top surface $z=N_{z,TI}$ (which is made in contact with the FMM later), the spin polarization $\langle k_{x},k_{y},n_{z}|{\boldsymbol\sigma}\otimes I|k_{x},k_{y},n_{z}\rangle$ of these surface states exhibits the spin-momentum locking\cite{Zhang09,Liu10,Yazyev10,Shan10,Zhang12}, as indicated by the red arrows that circulate along the Dirac cone in Fig.~\ref{fig:3DTIFMMslab_schematics} (b). Note that the bulk bands of the TI is also spin polarized, as indicated by the red arrows in Fig.~\ref{fig:3DTIFMMslab_schematics} (b) on the bands that are gapped.


The charge and spin current operators are constructed from the local charge and spin density
\begin{eqnarray}
n_{i}=\sum_{I\sigma}c_{iI\sigma}^{\dag}c_{iI\sigma},\;\;\;
m_{i}^{a}=\sum_{I}c_{iI\alpha}^{\dag}\sigma_{\alpha\beta}^{a}c_{iI\beta},
\end{eqnarray}
whose equations of motion can be written in the form of continuity equations 
\begin{eqnarray}
&&\dot{n}_{i}=\frac{i}{\hbar}\left[H,n_{i}\right]=-{\boldsymbol\nabla}\cdot{\bf J}_{i}^{0}=-\frac{1}{a}\sum_{\delta}\left(J_{i,i+\delta}^{0}+J_{i,i-\delta}^{0}\right),
\nonumber \\
&&\dot{m}_{i}^{a}=\frac{i}{\hbar}\left[H,m_{i}^{a}\right]=-{\boldsymbol\nabla}\cdot{\bf J}_{i}^{a}+\tau_{i}^{a}
\nonumber \\
&&=-\frac{1}{a}\sum_{\delta}\left(J_{i,i+\delta}^{a}+J_{i,i-\delta}^{a}\right)+\tau_{i}^{a},
\label{ndot_Mdot_commutator}
\end{eqnarray}
which defines the local charge and spin currents $J_{i,i+\delta}^{a}$ running from site $i$ to $i+\delta$, and $J_{i,i-\delta}^{a}$ that run from $i$ to $i-\delta$, and $\tau_{i}^{a}$ is the local torque that comes from the $J_{ex}$ term in Eq.~(\ref{3DTIFMM_Hamiltonian}). Their precise forms are detailed in Appendix \ref{apx:current_operators}. We will define a local charge and spin current by considering the current running along positive bonds in either $x$ or $y$ direction as a function of out-of-plane coordinate $z$
\begin{eqnarray}
J_{x}^{y}(z)\equiv J_{i,i+a}^{y},\;\;\;J_{y}^{x}(z)\equiv J_{i,i+b}^{x},
\end{eqnarray}
and investigate their profile due to proximity to the FMM.

\subsection{Percolation of topological surface states into the FMM \label{sec:percolation_surface_state}}



\begin{figure}[h]
\begin{center}
\includegraphics[clip=true,width=0.95\columnwidth]{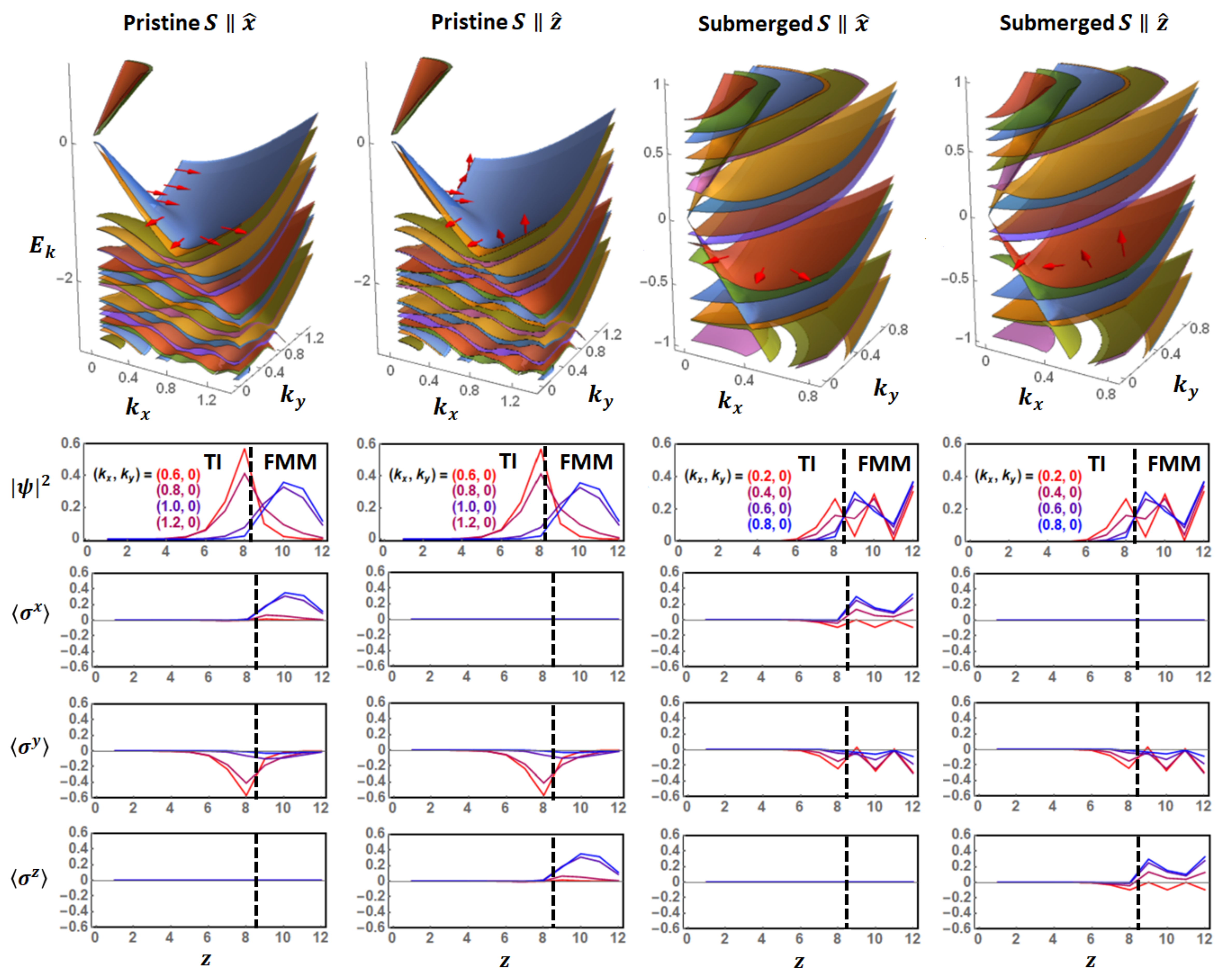}
\caption{The pristine and submerged types of band structures $E_{\bf k}$ that distinguishes whether the Dirac cone submerges into the FMM bulk bands, with magnetization pointing in-plane ${\bf S}\parallel{\hat{\bf x}}$ and out-of-plane ${\bf S}\parallel{\hat{\bf z}}$. We choose $N_{z,TI}=8$ layers of TI and $N_{z,FM}=4$ layers of FMM. Red arrows show the spin polarization of the eigenstate at several selected $(k_{x},k_{y},n_{z})$ that gradually moves from Dirac cone-like states to FMM bulk-like states. The bottom panels show the wave functions $|\psi|^{2}$ and the spin components $\langle\sigma^{a}\rangle$ as a function of out-of-plane coordinate $z$ for some of these $(k_{x},k_{y},n_{z})$. } 
\label{fig:bands_spin_SxSySz}
\end{center}
\end{figure}

Since the TI/FMM contact requires to align the work functions of the two materials, as that occurs in the semiconductor-metal junctions\cite{Cowley65,Tung14}, the FMM bands can be shifted relative to the TI bands. This shift in our lattice model is simulated by adjusting the FMM chemical potential $\mu_{F}$ in Eq.~(\ref{3DTIFMM_Hamiltonian}). As a result of the shift, there can be what we call the pristine type of band structure where the large part of the Dirac cone does not overlap with the FMM bulk bands, and the submerged type where the Dirac cone submerges deeply into the FMM bulk bands\cite{Zegarra20}. We choose the following $\mu_{F}$ to investigate these two generic types of band structure
\begin{eqnarray}
{\rm pristine:}\;\;\;\mu_{F}=0.5,\;\;\;\;\;
{\rm submerged:}\;\;\;\mu_{F}=-2.
\end{eqnarray}
Figure \ref{fig:bands_spin_SxSySz} shows the band structures, wave functions, and spin polarizations for the pristine and submerged types, with magnetization pointing in-plane ${\bf S}\parallel{\hat{\bf x}}$ and out-of-plane ${\bf S}\parallel{\hat{\bf z}}$, which reveal the following interesting features.


For the pristine type of band structures, as moving from small to large momentum, the Dirac cone gradually merges into the FMM bulk bands. The spin polarization of the eigenstate $|k_{x},k_{y},n_{z}\rangle$ gradually rotates from that given by the spin momentum locking of the Dirac cone to that along the magnetization, as indicated by the red arrows in the top panels of Fig.~\ref{fig:bands_spin_SxSySz}. The surface state wave function $|\psi|^{2}$ (localized at the TI boundary) gradually merges with the FMM quantum well state wave function (standing wave inside the FMM) as moving to large momentum. The spatial profile of the spin polarization of $|k_{x},k_{y},n_{z}\rangle$ is such that the wave function in the TI region remains largely polarized in the spin-momentum locking direction, with a small component parallel to the magnetization, whereas the wave function in the FMM region is mainly polarized along the magnetization.

For the submerged type of band structure, the Dirac cone overlaps and intertwines with the FMM bulk bands drastically. Tracking the states originating from the Dirac cone reveals that the surface state is even more hybridized with the FMM quantum well state, yielding a wave function that has the feature of both states, i.e., evanescent in the TI region and standing wave (possibly of higher harmonics\cite{Zegarra20}) in the FMM region. A significant spin polarization along the magnetization is induced in the TI region, indicating that the spin-momentum locking in the TI is distorted significantly. As we shall see in the following sections, these peculiar properties of percolated surface states have a profound influence on the spintronic properties of the TI/FMM bilayers.


\subsection{Persistent charge current \label{sec:persistent_charge_current}}

\begin{figure}[ht]
\begin{center}
\includegraphics[clip=true,width=0.99\columnwidth]{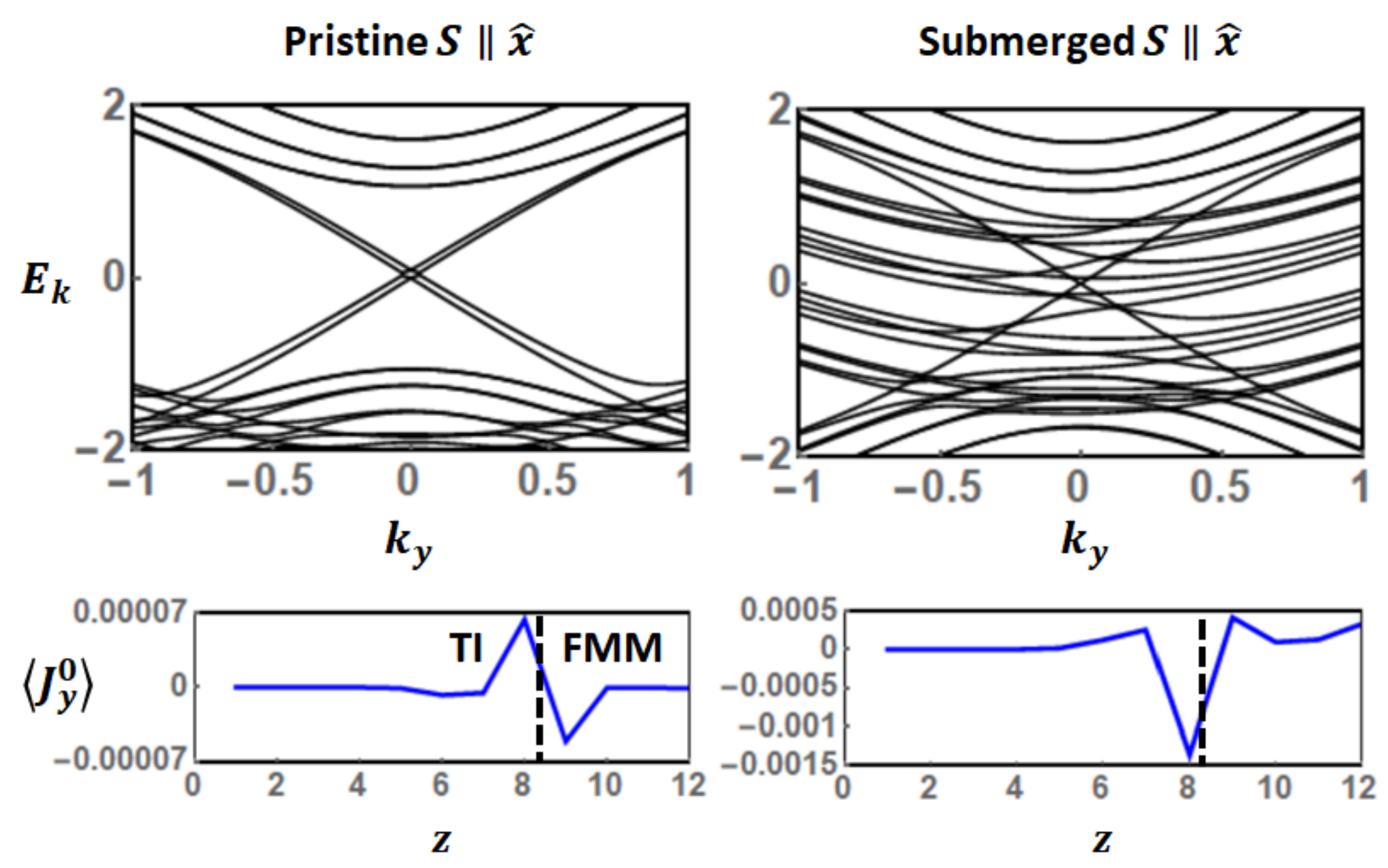}
\caption{(top) The pristine and submerged type of band structures as a function of $k_{y}$ at $k_{x}=0$, with magnetization pointing along ${\bf S}\parallel{\hat{\bf x}}$. The asymmetry of the band structure between $+k_{y}$ and $-k_{y}$ is clearly visible. (bottom) The corresponding persistent charge current $\langle J_{y}^{0}\rangle$ flowing along ${\hat{\bf y}}$ direction as a function of out-of-plane coordinate $z$. } 
\label{fig:Eky_J0y_Sx}
\end{center}
\end{figure}

The dispersion for the case of out-of-plane magnetization ${\bf S}\parallel{\hat{\bf z}}$ is symmetric among momenta $(\pm k_{x},\pm k_{y})$. However, if the magnetization lies in-plane, then the dispersion becomes asymmetric in the direction perpendicular to the magnetization. This is because in the profile of the spin-momentum locking, the states polarized along the magnetization becomes energetically more favorable than the states polarized in the opposite direction, hence tilting the whole band structure\cite{Manchon08,Manchon09,Gambardella11}. As an example, in Fig.~\ref{fig:Eky_J0y_Sx} the case of ${\bf S}\parallel{\hat{\bf x}}$ is present, which renders a dispersion asymmetric between $+k_{y}$ and $-k_{y}$ for either the pristine or the submerged type of band structures.

\onecolumngrid

\twocolumngrid

The asymmetric dispersion prompts us to investigate the possibility of an equilibrium persistent current in the system, since the dispersion seems to imply the electron motions in positive and negative directions are different. However, it is easy to see that the asymmetric dispersion does not yield a net current at equilibrium, or equivalently the Fermi sea does not carry a net group velocity. This is because the expectation value of the in-plane velocity operator $v_{a}=\left\{v_{x},v_{y}\right\}$ is simply the group velocity\cite{Nagaosa08}
\begin{eqnarray}
&&\langle u_{k_{x},k_{y},n_{z}}|v_{a}|u_{k_{x},k_{y},n_{z}}\rangle=\langle u_{k_{x},k_{y},n_{z}}|\frac{1}{\hbar}\frac{\partial H}{\partial k_{a}}|u_{k_{x},k_{y},n_{z}}\rangle
\nonumber \\
&&=\frac{\partial E(k_{x},k_{y},n_{z})}{\hbar\partial k_{a}}.
\end{eqnarray}
The expectation value integrated over momentum vanishes identically
\begin{eqnarray}
\langle v_{a}\rangle&=&\sum_{n_{z}}\int\frac{dk_{x}}{2\pi}\int\frac{dk_{y}}{2\pi}\frac{\partial E(k_{x},k_{y},n_{z})}{\hbar\partial k_{a}}f(E(k_{x},k_{y},n_{z}))
\nonumber \\
&=&0\;,
\label{ground_state_current}
\end{eqnarray}
where $f(E(k_{x},k_{y},n_{z}))=1/\left(e^{E(k_{x},k_{y},n_{z})/k_{B}T}+1\right)$ is the Fermi function. Thus there is no net charge current in either the direction parallel or perpendicular to the magnetization.

However, using the current operator in Appendix \ref{apx:current_operators}, we reveal that there exists an equilibrium {\it local} charge current flowing in the direction perpendicular to the magnetization. As shown in Fig.~\ref{fig:Eky_J0y_Sx} for the ${\bf S}\parallel{\hat{\bf x}}$ case, a laminar flow of persistent charge current $\langle J_{y}^{0}(z)\rangle\equiv \langle J_{i,i+b}^{0}\rangle$, meaning that the direction of flow is along $+{\hat{\bf y}}$ or $-{\hat{\bf y}}$ depends on the out-of-plane position $z$, is uncovered. The laminar current exists in both the TI region $z\leq N_{z,TI}$ and the FMM region $N_{z,TI}\leq z\leq N_{z,TI}+N_{z,FM}$, and sums to zero $\sum_{z}\langle J_{y}^{0}(z)\rangle\approx 0$ up to numerical precision, in agreement with Eq.~(\ref{ground_state_current}). This current is absent if the magnetization points entirely out-of-plane ${\bf S}\parallel{\hat{\bf z}}$, and there is no current along the direction parallel to the magnetization $\langle J_{x}^{0}(z)\rangle\equiv \langle J_{i,i+a}^{0}\rangle$, indicating the current indeed originates from the asymmetric band structure induced by the in-plane magnetization. The band structure origin makes this equilibrium current easily persist up to room temperature and macroscopic scale, which is an advantage compared to that induced at the topological superconductor/FMM interface\cite{Brydon13,Schnyder13}.

\subsection{Persistent spin current \label{sec:persistent_spin_current}}

The spin-momentum locking of the surface states shown in Fig.~\ref{fig:3DTIFMMslab_schematics} (b) has speculated a surface spin current at equilibrium\cite{Buttiker09,Sonin11,Ando13,Maekawa17}. For an isolated TI with OBC imposed in the ${\hat{\bf z}}$ direction, one expects the surface states to cause a spin current $\langle J_{y}^{x}\rangle$ polarized along ${\hat{\bf x}}$ and flowing along ${\hat{\bf y}}$, and a spin current $\langle J_{x}^{y}\rangle$ polarized along ${\hat{\bf y}}$ and flowing along ${\hat{\bf x}}$ of equal magnitude. The spin currents should be localized at the two surfaces $z=1$ and $z=N_{y,TI}$, and the directions of flow are opposite between the two surfaces. 

However, it is shown recently that the above naive picture of equilibrium surface spin current has a serious flaw, namely it does not take into account the contribution from the valence bands\cite{Chen20_absence_edge_current}. For the cubic lattice model of an isolated TI, i.e., the $i\in TI$ terms in Eq.~(\ref{3DTIFMM_Hamiltonian}), the surface spin current produced by the surface states is in fact canceled out exactly by the contribution from the valence bands, rendering no net surface spin current. This surprising statement is valid regardless the temperature and parameters within the cubic lattice model. A finite surface spin current appears only when the chemical potential is shifted away from the Dirac point, since the cancellation from the valence bands is not complete in this case. Thus a variety of mechanisms in reality that shift chemical potential locally or globally, such as doping\cite{Hsieh09,Zhang11,Kondou16}, gating, impurities\cite{Beidenkopf11}, and surface band bending\cite{Bahramy12}, can all be used to promote the surface spin current\cite{Chen20_absence_edge_current}.

\begin{figure}[ht]
\begin{center}
\includegraphics[clip=true,width=0.99\columnwidth]{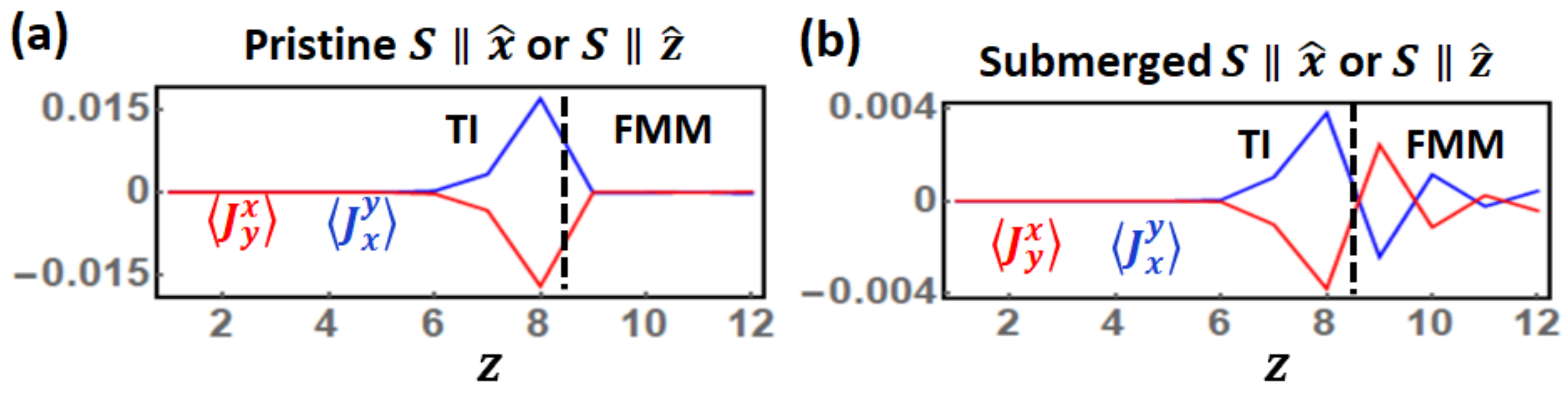}
\caption{(a) The spin current in the TI/FMM slab for the pristine type of band structure, and (b) for the submerged type. The spatial profile and magnitude of the spin current only varies by few percent as changing the direction of the magnetization ${\bf S}$. } 
\label{fig:spin_current_results}
\end{center}
\end{figure}


Using the lattice model in Eq.~(\ref{3DTIFMM_Hamiltonian}), we further uncover that an equilibrium spin current occurs when the TI is made in contact with the FMM, even if the Dirac point resides at the chemical potential. The spin current in the TI/FMM bilayers shown in Fig.~\ref{fig:spin_current_results} (b) and (c) has the following features: (i) For the pristine type of band structure, the spin current mainly concentrates in the TI region near the interface, but for the submerged case the spin current in the FMM region is dramatically enhanced. (ii) The sptial profile and magnitude of the spin current remain roughly the same for any direction of magnetization ${\bf S}$, with only few percent variation. (iii) The relation $\langle J_{y}^{x}\rangle=-\langle J_{x}^{y}\rangle$ is satisfied for the out-of-plane magnetization case ${\bf S}\parallel{\hat{\bf z}}$, whereas for all other magnetization directions they are approximately equal $\langle J_{y}^{x}\rangle\approx-\langle J_{x}^{y}\rangle$. (iv)  The spin current is also a laminar flow whose direction of flow depends on the out-of-plane position $z$, which is is particularly evident for the submerged type of band structure shown in Fig.~\ref{fig:spin_current_results} (c), and the spin current does not sum to zero, i.e., there is a net spin current.


\subsection{Linear response theory for the magnetoelectric susceptibility \label{sec:linear_response_chi}}

The current-induced spin torque originates from the nonequilibrium spin accumulation in the FMM caused by a bias voltage. In this section, we aim to calculate such a nonequilibrium response (in contrast to the equilibrium charge and spin currents in Secs.~\ref{sec:persistent_charge_current} and \ref{sec:persistent_spin_current}). Our goal is to calculate the local spin accumulation $\sigma^{b}(i,t)$ induced by a perturbation $H'(t')$ in the Hamiltonian by means of a linear response theory\cite{Zegarra20,Chen09_resistivity_upturn,Takigawa02}
\begin{eqnarray}
\sigma^{b}(i,t)=-i\int_{-\infty}^{t}dt'\langle\left[\sigma^{b}(i,t),H'(t')\right]\rangle\;,
\label{sigma_linear_response}
\end{eqnarray}
where $\sigma^{b}(i,t)=\sum_{I\beta\gamma}c_{iI\beta}^{\dag}(t)\sigma^{b}_{\beta\gamma}c_{iI\gamma}(t)$ is the $b=\left\{x,y,z\right\}$ component of the spin operator at position $i$, and $c_{iI\gamma}(t)$ are the electron operators defined in the Heisenberg picture. The time-variation of the longitudinal component of the vector field $A(j,t')=A(j)e^{-i\omega t'}$ induces the electric field $E=-\frac{\partial A}{\partial t}=i\omega A$ along ${\hat{\bf x}}$ direction and the electric current, as the situation in the experimental setup, and hence the perturbation is
\begin{eqnarray}
H'(t')=-\sum_{j}J_{x}^{0}(j,t')A(j,t')\;,
\end{eqnarray}
where we have abbriviated the current operator flowing in the ${\hat{\bf x}}$ direction by $J_{x}^{0}(j,t')\equiv J_{j,j+a}^{0}(t')$ in comparison with the lattice notation in Eq.~(\ref{J0_operators}). Consequently, the commutator in Eq.~(\ref{sigma_linear_response}) reads 
\begin{eqnarray}
\left[\sigma^{b}(i,t),H'(t')\right]=\frac{i}{\omega}\sum_{j}e^{i\omega(t-t')}E(j,t)\left[\sigma^{b}(i,t),J_{x}^{0}(j,t')\right]\;,
\nonumber \\
\end{eqnarray}
where $E(i,t)=E^{0}e^{i{\bf q\cdot r}_{i}-i\omega t}$. The local spin accumulation in Eq.~(\ref{sigma_linear_response}) then becomes 
\begin{eqnarray}
&&\sigma^{b}({\bf r},t)
\nonumber \\
&=&\sum_{j}\int_{-\infty}^{\infty}dt'e^{i\omega(t-t')}\frac{1}{\omega}\theta(t-t')
\langle\left[\sigma^{b}(i,t),J_{x}^{0}(j,t')\right]\rangle E(j,t)
\nonumber \\
&=&\sum_{j}\int_{-\infty}^{\infty}dt'e^{i\omega(t-t')}\frac{i\pi^{b}(i,j,t-t')}{\omega} E(j,t)
\nonumber \\
&=&\sum_{j}\frac{i\pi^{b}(i,j,\omega)}{\omega} E(j,t)\equiv \sum_{j}\chi^{b}(i,j,\omega) E(j,t)\;.
\label{sigmab_chiij_nonlocal}
\end{eqnarray}
Here $\chi^{b}(i,j,\omega)$ is the response coefficient for the contribution to the $\sigma^{b}(i,t)$ at site $i$ due to the longitudinal electric field $E(j,t)$ applied at site $j$. Assuming the electric field is constant everywhere ${\bf q}\rightarrow 0$ such that $E(i,t)=E(j,t)=E^{x}e^{-i\omega t}$, Eq.~(\ref{sigmab_chiij_nonlocal}) may be written in a form that defines the magnetoelectric susceptibility
\begin{eqnarray}
\sigma^{b}(i,t)=\left\{\sum_{j}\chi^{b}(i,j,\omega)\right\} E(i,t)=\chi^{b}(i,\omega)E(i,t)\;,
\label{sigma_chi_E}
\nonumber \\
\end{eqnarray} 
The real part of the DC magnetoelectric susceptibility is what we aim to calculate
\begin{eqnarray}
\lim_{\omega\rightarrow 0}{\rm Re}\chi^{b}(i,\omega)=\lim_{\omega\rightarrow 0}{\rm Re}\left\{\frac{i}{\omega}\sum_{j}\pi^{b}(i,j,\omega)\right\}\;,
\label{Rechi_DC_limit}
\end{eqnarray}
After diagonalizing the lattice Hamiltonian in Eq.~(\ref{3DTIFMM_Hamiltonian}), we obtain the eigenstate $|n\rangle$ with eigenenergy $E_{n}$, and calculate the retarded response function $\pi^{b}(i,j,\omega)$ by\cite{Zegarra20,Chen09_resistivity_upturn,Takigawa02} 
\begin{eqnarray}
\pi^{b}(i,j,\omega)=\sum_{m,n}\langle n|\sigma^{b}(i)|m\rangle\langle m|J_{x}^{0}(j)|n\rangle\frac{f(E_{n})-f(E_{m})}{\omega+E_{n}-E_{m}+i\eta},
\nonumber \\
\label{pibij_formula}
\end{eqnarray}
where $\eta$ is a small artificial broadening. We are lead to 
\begin{eqnarray}
&&\lim_{\omega\rightarrow 0}{\rm Re}\chi^{b}(i,\omega)
\nonumber \\
&&=-\sum_{j}\sum_{m,n}\langle n|\sigma^{b}(i)|m\rangle\langle m|J_{x}^{0}(j)|n\rangle\tilde{F}(E_{n},E_{m})\;,
\nonumber \\
&&\tilde{F}(E_{n},E_{m})=\int d\omega\,\frac{\eta}{(\omega-E_{n})^{2}+\eta^{2}}\left(\frac{1}{\pi}\frac{\partial f(\omega)}{\partial \omega}\right)
\nonumber \\
&&\;\;\;\;\;\;\;\;\;\;\;\;\;\;\times\frac{\eta}{(\omega-E_{m})^{2}+\eta^{2}}\;.
\label{chi_Fnm_formula}
\end{eqnarray}
Numerically, including about $\sim 100$ states near the Fermi surface in the summation $\sum_{n}$ and $\sum_{m}$ is already sufficient to obtain a precise $\chi^{b}$, since the nonequilibrium magnetoelectric response is mainly contributed from these states, and we choose the artificial broadening $\eta=0.05$ (mean free time $\tau\sim 10^{-14}$s). Note that the diagonal elements vanish $\tilde{F}(E_{n},E_{n})=0$ as implied in the definition in Eq.~(\ref{pibij_formula}).

The following subtleties must be taken care of when applying the above linear response theory to our lattice model in Eq.~(\ref{3DTIFMM_Hamiltonian}). For an isolated TI, Kramers theorem dictates that every eigenstate is two-fold spin degenerate. Moreover, the surface states localized at the top $z=N_{z,TI}$ and bottom $z=1$ surfaces are degenerate, in addition to the degeneracy caused by various spatial symmetries of the cubic lattice. The wave functions that are degenerate can arbitrarily mix up in our numerical calculation, which complicates the evaluation of the matrix elements $\langle n|\sigma^{b}(i)|m\rangle$ and $\langle m|J_{x}^{0}(j)|n\rangle$ in Eq.~(\ref{chi_Fnm_formula}). Thus the following treatments must be implimented to obtain a reasonable magnetoelectric response. Firstly, we consider the TI/FMM bilayer instead of an isolated TI, such that the coupling $t_{B}\neq 0$ to the FMM on the top surface removes the degeneracy between the two surfaces. Despite this coupling, the magnetoelectric susceptibility at the bottom surface $y=1$ still accurately captures the Edelstein effect of an isolated TI. Secondly, we add random point-like impurities into the lattice
\begin{eqnarray}
H_{imp}=U_{imp}\sum_{i\in imp,I\sigma}c_{iI\sigma}^{\dag}c_{iI\sigma},
\end{eqnarray}
where $i\in imp$ denotes the impurity sites. We consider a relatively high impurity density $10\%$ for the sake of removing spatial degeneracies and smearing out the energy spectrum, such that the accuracy of the numerical calculation can be improved. With these treatments, we estimate that our numerical calculation can reach about $70\%\sim 80\%$ accuracy, which is sufficient to draw conclusions.

\begin{figure}[ht]
\begin{center}
\includegraphics[clip=true,width=0.99\columnwidth]{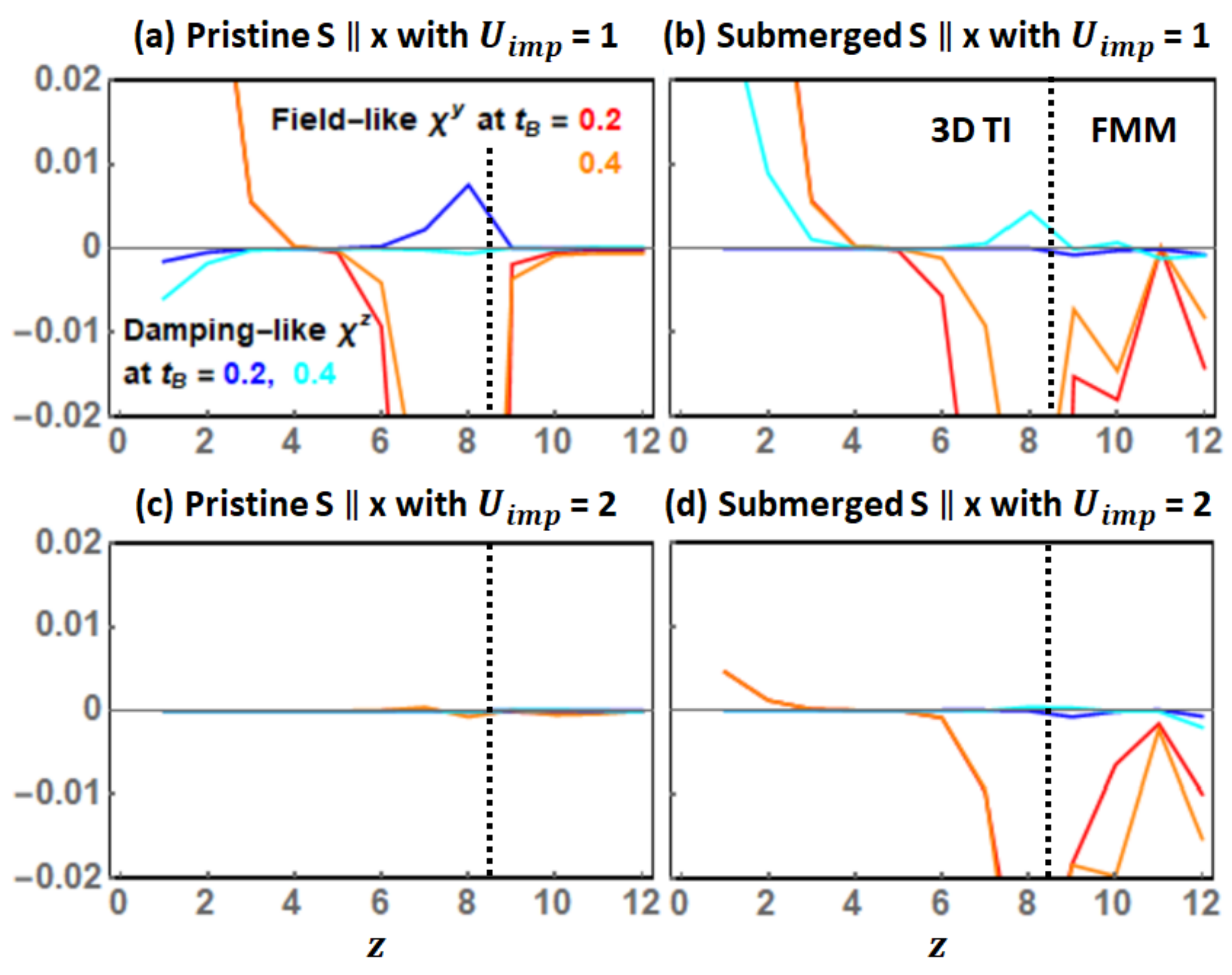}
\caption{The field-like $\chi^{y}$ and damping-like $\chi^{z}$ magnetoelectric susceptibility induced by an external electric field along ${\hat{\bf x}}$ direction and a magnetization also in the same direction ${\bf S}\parallel{\hat{\bf x}}$, averaged over planar coordinates $(x,y)$ and then plotted as a function of out-of-plane coordinate $z$. The four panels correspond to the two different types of band structures in Fig.~\ref{fig:bands_spin_SxSySz} labeled by pristine and submerged, and at $10\%$ of impurities with two different impurity potentials $U_{imp}=1$ and $2$. } 
\label{fig:chi_Sx_Uimp1_Uimp2}
\end{center}
\end{figure}

\subsubsection{Magnetization direction ${\bf S}\parallel{\hat{\bf x}}$}

The result of the simulation for the magnetization pointing along the current direction ${\bf S}\parallel{\hat{\bf x}}$ is shown in Fig.~\ref{fig:chi_Sx_Uimp1_Uimp2}, where the magnetoelectric susceptibility $\chi^{b}$ averaged over the planar directions $(x,y)$ plotted as a function of out-of-plane coordinate $z$ is presented for the pristine and submerged types of band structures, at two different values of interface hopping $t_{B}=0.2$ and $0.4$. We consider two different impurity potentials $U_{imp}=1$ and $2$. To interpret these results, note that for an isolated 3D TI, the spin-momentum locking of the surface states shown in Fig.~\ref{fig:3DTIFMMslab_schematics} (b) is expected to give a current-induced spin accumulation polarized along ${\hat{\bf y}}$ at the $z=1$ surface and $-{\hat{\bf y}}$ at the $z=N_{z,TI}$ surface, i.e., the Edelstein effect, which is correctly captured by the large $\chi^{y}$ in Fig.~\ref{fig:chi_Sx_Uimp1_Uimp2} (red and orange lines). Comparing the data at different impurity potentials, one sees that $\chi^{y}$ at the free surface $z=1$ is dramatically reduced at large impurity potential $U_{imp}=2$. This is qualitatively consistent with a recent analysis of the Edelstein effect based on a semiclassical approach\cite{Chen20_TI_Edelstein}, which suggests that the current-induced spin polarization reduces quadratically with the impurity potential $\chi^{y}(z=1)\propto 1/U_{imp}^{2}$. The absolute magnitude of $\chi^{y}(z=1)$ at $U_{imp}=1$ is the numerical number $\chi^{y}(z=1)\sim 0.1$ multiplied by $ae/t\sim 10^{-9}$mC/J. At the typical experimental charge current $j_{c}\sim 10^{11}$A/m$^{2}$ and the electrical conductivity of the FMM $\sim 10^{7}$S/m, the corresponding electric field is $E\sim 10^{4}$kgm/C$s^{2}$, which according to Eq.~(\ref{sigma_chi_E}) yields a spin polarization per unit cell $\sigma^{b}(i)\sim 10^{-6}$ in units of Bohr magneton.

Near the TI/FMM interface, from Fig.~\ref{fig:chi_Sx_Uimp1_Uimp2} one sees that $\chi^{b}$ extends into the FMM at $z\geq 9$. Because an isolated FMM has $\chi^{b}=0$ everywhere (assuming no other mechanisms give the spin accumulation, such as Rashba spin-orbit coupling), the finite $\chi^{b}$ in the FMM entirely comes from the proximity to the TI. Moreover, from Eq.~(\ref{chi_Fnm_formula}) one sees that $\chi^{b}$ originates from the states near the chemical potential $E_{\bf k}=0$, which include both the surface state Dirac cone and the FMM bulk bands according to the band structures in Figs.~\ref{fig:bands_spin_SxSySz} and \ref{fig:Eky_J0y_Sx}. The spin torque $d{\bf S}/dt$ on the magnetization is given by the averaged spin accumulation in the FMM region
\begin{eqnarray}
\frac{d{\bf S}}{dt}=\frac{J_{ex}}{\hbar}\left[\frac{1}{N_{z,FM}}\sum_{z\in FM}{\boldsymbol\chi}(z)E^{x}\right]\times {\bf S},
\label{Landau_Lifshitz}
\end{eqnarray}
following the usual Landau-Lifshitz dynamics. Because the Edelstein effect of an isolated TI gives a spin accumulation polarized along ${\hat{\bf y}}$, it is customary to define the field-like torque to be along ${\bf S}\times{\hat{\bf y}}$ and the damping-like torque to be along ${\bf S}\times({\bf S}\times{\hat{\bf y}})$. From Fig.~\ref{fig:chi_Sx_Uimp1_Uimp2}, it follows that the dominate component is the field-like $\chi^{y}$ (red and orange lines), and the damping-like $\chi^{z}$ (blue and light blue lines) is generally one order of magnitude smaller. In addition, both components are much larger in the submerged type of band structure, and moreover the spatial profile of $\chi^{y}$ resembles the wave function profile $|\psi|^{2}$ in Fig.~\ref{fig:bands_spin_SxSySz}, suggesting that the percolation of the surface state is crucial to the magnitude of the spin torque.

This predominantly field-like torque is similar to that occurs in the 2D version of this problem\cite{Zegarra20}, which has been attributed to the real wave functions of the percolated surface states that cannot accumulate a spin-dependent phase, unlike the spin-transfer torque in usual metallic heterostructures\cite{Berger96,Slonczewski96} and spin Hall systems\cite{Chen15_spin_transfer_torque,Sakanashi18} where the spin polarized plane waves accumulates a spin-dependent phase that eventually yields a damping-like torque. At a typical external electric current $j_{c}\sim 10^{11}$A/m$^{2}$, the spin polarization is basically the numerical values of $\chi^{b}$ multiplied by GHz, which is close to that observed experimentally\cite{Mellnik14}.

\begin{figure}[ht]
\begin{center}
\includegraphics[clip=true,width=0.99\columnwidth]{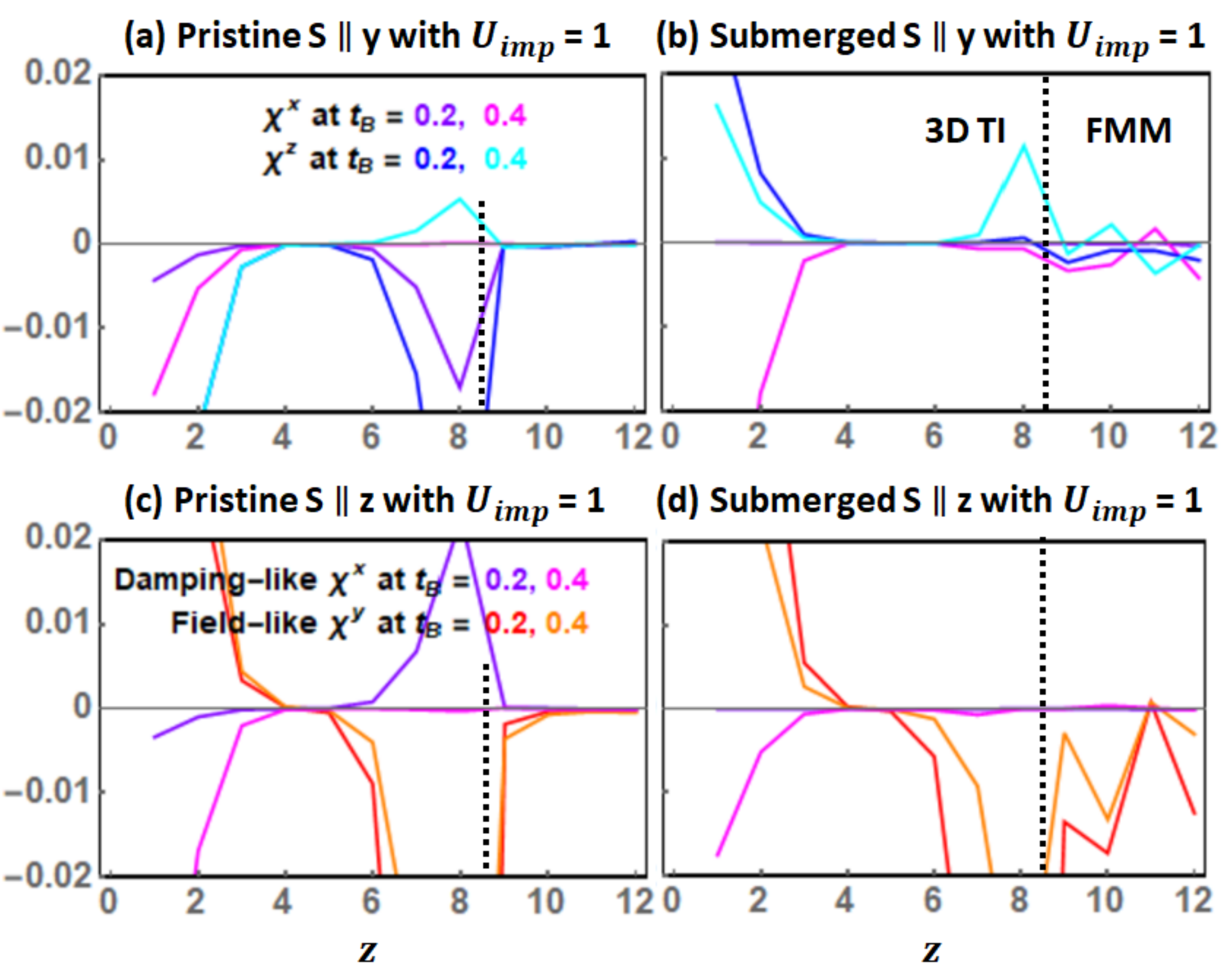}
\caption{The planar averaged magnetoelectric susceptibility $\chi^{b}$ at magnetization direction ${\bf S}\parallel{\hat{\bf y}}$ and ${\bf S}\parallel{\hat{\bf z}}$  plotted as a function of out-of-plane coordinate $z$, for the pristine and submerged types of band structures. The impurity potential is fixed at $U_{imp}=1$. } 
\label{fig:chi_Sy_Sz_Uimp1}
\end{center}
\end{figure}

\subsubsection{Magnetization directions ${\bf S}\parallel{\hat{\bf y}}$ and ${\bf S}\parallel{\hat{\bf z}}$}

Figure \ref{fig:chi_Sy_Sz_Uimp1} (a) and (b) show the result for the magnetization along ${\bf S}\parallel{\hat{\bf y}}$. Focusing on the free surface $z=1$, we uncover that the magnetoelectric susceptibility is not only polarized in the direction $\chi^{y}$ (not shown) expected from the Edelstein effect, but also has $\chi^{x}$ and $\chi^{z}$ components. As these two components are beyond the usual semiclassical picture that treats each impurity as an independent scatterer\cite{Chen20_TI_Edelstein}, they are attributed to the interference effect at high impurity densities ($10\%$ in our numerical calculation), and their magnitudes are generally few times or one order smaller than $\chi^{y}$. Moreover, although this ${\bf S}\parallel{\hat{\bf y}}$ case is not expected to produce any torque according to the discussion after Eq.~(\ref{Landau_Lifshitz}), the magnetization in the FMM in fact experiences a torque in both ${\hat{\bf x}}$ and ${\hat{\bf z}}$ directions as a result of this interference effect and the percolation of the surface state. Once again $\chi^{x}$ and $\chi^{z}$ are larger in the submerged type of band structures, and has a spatial profile that varies significantly with the interface hopping $t_{B}$.

Finally, we present the result for the out-of-plane magnetization ${\bf S}\parallel{\hat{\bf z}}$ in Fig.~\ref{fig:chi_Sy_Sz_Uimp1} (c) and (d). This case is similar to the other two magnetization directions, namely we observe a predominantly field-like spin torque due to the $\chi^{y}$ component whose percolation into the FMM is more prominent in the submerged type of band structure. The damping-like component $\chi^{x}$ is rather insignificant compared to the field-like component. Nevertheless, at the free surface $z=1$ of the TI a significant amount of $\chi^{x}$ is induced due to the impurity effect. The magnitude of all these components are reduced at larger impurity potential $U_{imp}$.


\section{Conclusions}

In summary, the spintronic properties of TI/FMM bilayers are investigated by means of a regularized cubic lattice model that simultaneously takes into account the surface state Dirac cone and the FMM bulk bands. We distinguish the pristine and the submerged types of band structures according to whether the Dirac cone overlaps with the FMM bulk bands, which is determined by the work functions of the two materials. Through investigating the wave function and spin polarization of the eigenstates at different momenta, we find that the surface state of the TI percolates into the FMM, and the spin polarization profile of the surface state is highly influenced by the magnetization of the FMM. In other words, the spin-momentum locking of the surface state is distorted by the magnetization. As moving from small to large momentum, the Dirac cone gradually merges with the FMM bulk bands, and the spin polarization gradually rotates to be along the magnetization. For the submerged type of band structure, the Dirac cone and the FMM bulk bands become highly intertwined, and hence it is rather ambiguous to distinguish the surface states and the FMM quantum well states.

Particularly for the case of in-plane magnetization, the combined effect of spin momentum locking and the coupling to the magnetization renders a band structure that is asymmetric in the direction perpendicular to the magnetization. As a result, the system develops a persistent laminar current whose direction of flow depends on the out-of-plane coordinate, and exists in both the TI layer and the FMM layer. This laminar persistent current paves a way for a magnetization induced room temperature persistent current that extends over macroscopic scale. Moreover, the proximity to the FMM also induces a laminar spin current flowing in both the TI and the FMM, whose spatial profile is roughly independent from the direction of magnetization, but highly influenced by the detail of the band structure. Finally, in the absence of interface Rashba spin-orbit coupling, the current-induced spin torque is contributed from both the Dirac cone and the FMM bulk bands, and is predominantly field-like along ${\bf S}\times{\hat{\bf y}}$ owing to the real wave functions of the percolated surface states, with a magnitude highly influenced by the degree of the percolation of the surface states. On the other hand, impurities can alter the spin accumulation caused by the surface state and generate a damping-like torque along ${\bf S}\times({\bf S}\times{\hat{\bf y}})$ in the FMM, and moreover cause a torque even if the magnetization points along ${\bf S}\parallel{\hat{\bf y}}$. We anticipate that these results can be verified experimentally by comparing samples with different impurity densities and band structures, and help to engineer the spin torque in these bilayers to suit proper applications.

The author acknowledges fruitful discussions with J. C. Egues, A. Zegarra, R. B. Muniz, and C. Lewenkopf, and the financial support from the productivity
in research fellowship of CNPq.



\appendix

\section{Detail of the charge and spin current operators \label{apx:current_operators}}

In practice, we may simplify the calculation of the current operators by the following method. Since only hopping terms in Eq.~(\ref{3DTIFMM_Hamiltonian}) contribute to the current operator, we focus on these terms that generally take the form
\begin{eqnarray}
H_{L\alpha M\beta}^{\delta}=\sum_{j}T_{L\alpha M\beta}^{\delta}c_{jL\alpha}^{\dag}c_{j+\delta M\beta}+T_{L\alpha M\beta}^{\delta\ast}c_{j+\delta M\beta}^{\dag}c_{jL\alpha}\;,
\nonumber \\
\label{HLalphaMbeta_delta}
\end{eqnarray}
which describes the hopping of electron between site/orbital/spin $jL\alpha$ and $j+\delta M\beta$ along the planar directions $\delta=\left\{a,b\right\}$, with $T_{L\alpha M\beta}^{\delta}$ the hopping amplitude. The hopping part of the total Hamiltonian is the summation of $H_{t}=\sum_{\delta}\sum_{L\alpha M\beta}H_{L\alpha M\beta}^{\delta}$. Directly evaluating the commutator and then comparing with the definitions in Eq.~(\ref{ndot_Mdot_commutator}), and separating the $i+\delta$ and $i-\delta$ parts yield
\begin{eqnarray}
J_{i,i+\delta}^{0}&=&\frac{ia}{\hbar}\sum_{IM}\left\{T_{I\sigma M\beta}^{\delta}c_{iI\sigma}^{\dag}c_{i+\delta M\beta}-T_{I\sigma M\beta}^{\delta\ast}c_{i+\delta M\beta}^{\dag}c_{iI\sigma}\right\},
\nonumber \\
J_{i,i+\delta}^{a}&=&\frac{ia}{\hbar}\sum_{IM}\left\{T_{I\lambda M\beta}^{\delta}c_{iI\eta}^{\dag}\sigma_{\eta\lambda}^{a}c_{i+\delta M\beta}\right.
\nonumber \\
&&\left.-T_{I\eta M\beta}^{\delta\ast}c_{i+\delta M\beta}^{\dag}\sigma_{\eta\lambda}^{a}c_{iI\lambda}\right\},
\end{eqnarray}
Collecting all the hopping terms, we arrive at the charge currents flowing the positive directions
\begin{eqnarray}
&&J_{i,i+a}^{0}=\frac{ia}{\hbar}\left.\left\{t_{\parallel}\sum_{I}\left[
c_{iI\uparrow}^{\dag}c_{i+a\overline{I}\downarrow}
-c_{iI\downarrow}^{\dag}c_{i+a\overline{I}\uparrow}\right]-h.c.\right\}\right|_{i\in TI}
\nonumber \\
&&+\frac{ia}{\hbar}\left.\left\{M_{2}\sum_{\sigma}\left[
-c_{is\sigma}^{\dag}c_{i+as\sigma}+c_{ip\sigma}^{\dag}c_{i+ap\sigma}\right]-h.c.\right\}\right|_{i\in TI}
\nonumber \\
&&-\frac{ia}{\hbar}t_{F}\left.\sum_{I\sigma}\left\{c_{iI\sigma}^{\dag}c_{i+aI\sigma}
-h.c.\right\}\right|_{i\in FM},
\nonumber \\
&&J_{i,i+b}^{0}=\frac{ia}{\hbar}\left.\left\{it_{\parallel}\sum_{I}\left[
-c_{iI\uparrow}^{\dag}c_{i+b\overline{I}\downarrow}
-c_{iI\downarrow}^{\dag}c_{i+b\overline{I}\uparrow}\right]-h.c.\right\}\right|_{i\in TI}
\nonumber \\
&&+\frac{ia}{\hbar}\left.\left\{M_{2}\sum_{\sigma}\left[
-c_{is\sigma}^{\dag}c_{i+bs\sigma}+c_{ip\sigma}^{\dag}c_{i+bp\sigma}\right]-h.c.\right\}\right|_{i\in TI}
\nonumber \\
&&-\frac{ia}{\hbar}t_{F}\left.\sum_{I\sigma}\left\{c_{iI\sigma}^{\dag}c_{i+bI\sigma}
-h.c.\right\}\right|_{i\in FM},
\label{J0_operators}
\end{eqnarray}
The spin currents polarized along $\sigma^{x}$ and flowing along positive directions are
\begin{eqnarray}
&&J_{i,i+a}^{x}=\frac{ia}{\hbar}\left.\left\{t_{\parallel}\sum_{I}\left[
-c_{iI\uparrow}^{\dag}c_{i+a\overline{I}\uparrow}
+c_{iI\downarrow}^{\dag}c_{i+a\overline{I}\downarrow}\right]-h.c.\right\}\right|_{i\in TI}
\nonumber \\
&&+\frac{ia}{\hbar}\left.\left\{M_{2}\sum_{\sigma}\left[
-c_{is\sigma}^{\dag}c_{i+as\overline{\sigma}}+c_{ip\sigma}^{\dag}c_{i+ap\overline{\sigma}}\right]-h.c.\right\}\right|_{i\in TI},
\nonumber \\
&&-\frac{ia}{\hbar}t_{F}\left.\sum_{I}\left\{c_{iI\alpha}^{\dag}\sigma_{\alpha\beta}^{x}c_{i+aI\beta}
-h.c.\right\}\right|_{i\in FM},
\nonumber \\
&&J_{i,i+b}^{x}=\frac{ia}{\hbar}\left.\left\{it_{\parallel}\sum_{I}\left[
-c_{iI\uparrow}^{\dag}c_{i+b\overline{I}\uparrow}
-c_{iI\downarrow}^{\dag}c_{i+b\overline{I}\downarrow}\right]-h.c.\right\}\right|_{i\in TI}
\nonumber \\
&&+\frac{ia}{\hbar}\left.\left\{M_{2}\sum_{\sigma}\left[
-c_{is\sigma}^{\dag}c_{i+bs\overline{\sigma}}
+c_{ip\sigma}^{\dag}c_{i+bp\overline{\sigma}}\right]-h.c.\right\}\right|_{i\in TI}
\nonumber \\
&&-\frac{ia}{\hbar}t_{F}\left.\sum_{I}\left\{c_{iI\alpha}^{\dag}\sigma_{\alpha\beta}^{x}c_{i+bI\beta}
-h.c.\right\}\right|_{i\in FM},
\end{eqnarray}
The spin currents polarized along $\sigma^{y}$ and flowing along positive directions are
\begin{eqnarray}
&&J_{i,i+a}^{y}=\frac{ia}{\hbar}\left.\left\{it_{\parallel}\sum_{I}\left[
c_{iI\uparrow}^{\dag}c_{i+a\overline{I}\uparrow}
+c_{iI\downarrow}^{\dag}c_{i+a\overline{I}\downarrow}\right]-h.c.\right\}\right|_{i\in TI}
\nonumber \\
&&+\frac{ia}{\hbar}\left.\left\{iM_{2}\sum_{\sigma}\left[
\sigma c_{is\sigma}^{\dag}c_{i+as\overline{\sigma}}-\sigma c_{ip\sigma}^{\dag}c_{i+ap\overline{\sigma}}\right]-h.c.\right\}\right|_{i\in TI}
\nonumber \\
&&-\frac{ia}{\hbar}t_{F}\left.\sum_{I}\left\{c_{iI\alpha}^{\dag}\sigma_{\alpha\beta}^{y}c_{i+aI\beta}
-h.c.\right\}\right|_{i\in FM},
\nonumber \\
&&J_{i,i+b}^{y}=\frac{ia}{\hbar}\left.\left\{t_{\parallel}\sum_{I}\left[
-c_{iI\uparrow}^{\dag}c_{i+b\overline{I}\uparrow}
+c_{iI\downarrow}^{\dag}c_{i+b\overline{I}\downarrow}\right]-h.c.\right\}\right|_{i\in TI}
\nonumber \\
&&+\frac{ia}{\hbar}\left.\left\{iM_{2}\sum_{\sigma}\left[
\sigma c_{is\sigma}^{\dag}c_{i+bs\overline{\sigma}}-\sigma c_{ip\sigma}^{\dag}c_{i+bp\overline{\sigma}}\right]-h.c.\right\}\right|_{i\in TI}
\nonumber \\
&&-\frac{ia}{\hbar}t_{F}\left.\sum_{I}\left\{c_{iI\alpha}^{\dag}\sigma_{\alpha\beta}^{y}c_{i+bI\beta}
-h.c.\right\}\right|_{i\in FM},
\end{eqnarray}
where $\sigma=\left\{\uparrow,\downarrow\right\}=\left\{+,-\right\}$. Finally, the spin currents polarized along $\sigma^{z}$ and flowing along positive directions are
\begin{eqnarray}
&&J_{i,i+a}^{z}=\frac{ia}{\hbar}\left.\left\{t_{\perp}\sum_{I}\left[
c_{iI\uparrow}^{\dag}c_{i+a\overline{I}\downarrow}
+c_{iI\downarrow}^{\dag}c_{i+a\overline{I}\uparrow}\right]-h.c.\right\}\right|_{i\in TI}
\nonumber \\
&&+\frac{ia}{\hbar}\left.\left\{M_{2}\sum_{\sigma}\left[
-\sigma c_{is\sigma}^{\dag}c_{i+as\sigma}+\sigma c_{ip\sigma}^{\dag}c_{i+ap\sigma}\right]-h.c.\right\}\right|_{i\in TI}
\nonumber \\
&&-\frac{ia}{\hbar}t_{F}\left.\sum_{I}\left\{c_{iI\alpha}^{\dag}\sigma_{\alpha\beta}^{z}c_{i+aI\beta}
-h.c.\right\}\right|_{i\in FM},
\nonumber \\
&&J_{i,i+b}^{z}=\frac{ia}{\hbar}\left.\left\{it_{\perp}\sum_{I}\left[
-c_{iI\uparrow}^{\dag}c_{i+b\overline{I}\downarrow}
+c_{iI\downarrow}^{\dag}c_{i+b\overline{I}\uparrow}\right]-h.c.\right\}\right|_{i\in TI}
\nonumber \\
&&+\frac{ia}{\hbar}\left.\left\{M_{2}\sum_{\sigma}\left[
-\sigma c_{is\sigma}^{\dag}c_{i+bs\sigma}+\sigma c_{ip\sigma}^{\dag}c_{i+bp\sigma}\right]-h.c.\right\}\right|_{i\in TI}
\nonumber \\
&&-\frac{ia}{\hbar}t_{F}\left.\sum_{I}\left\{c_{iI\alpha}^{\dag}\sigma_{\alpha\beta}^{z}c_{i+bI\beta}
-h.c.\right\}\right|_{i\in FM}.
\end{eqnarray}
The expectation values of these current operators can then be evaluated using the eigenstates $|n\rangle$ after diagonalizing the lattice Hamiltonian. 
Alternatively, one may perform the Fourier transform in Eq.~(\ref{Fourier_transform_kxky}) and then evaluate the expectation values of these current operators in the $(k_{x},k_{y},n_{z})$ basis.

\bibliography{Literatur}

\end{document}